\documentclass[10pt,conference]{IEEEtran}

\usepackage[T1]{fontenc}
\usepackage{cite}
\usepackage{amsmath,amsfonts,amsxtra,amssymb,latexsym,amscd,amsthm,mathrsfs,bm}

\usepackage{graphicx}

\usepackage{xcolor}
\usepackage{algorithm}
\usepackage{algorithmic}
\usepackage[small]{caption}
\usepackage{subcaption}
\usepackage{comment}
\usepackage{enumitem} 
\usepackage{epstopdf}					
\newcommand{\E}{\mathbb{E}}

\begin{document}
\title{Analysis and Compensation of Receiver IQ Imbalance and Residual CFO Error for AFDM
\author{\IEEEauthorblockN{Nimesha Gunasekara and Ebrahim Bedeer}
	\IEEEauthorblockA{Department of Electrical and Computer Engineering, University of Saskatchewan, 		Saskatoon, SK, Canada\\
		Emails: jgx691@mail.usask.ca and e.bedeer@usask.ca
}
}}

\maketitle

\begin{abstract}
Affine frequency division multiplexing (AFDM) is a promising waveform  for future wireless communication systems. 
In this paper, we analyze the impact of receiver in-phase and quadrature  (IQ) imbalance and residual carrier frequency offset (CFO) error on AFDM signals. 
Our analysis shows that the receiver IQ imbalance may not preserve the sparsity of the AFDM effective channel matrix because of the complex-conjugate operator of the discrete affine Fourier transform (DAFT). Moreover, the residual CFO error causes energy leakage in the effective channel matrix in the affine domain.
To mitigate these effects, we extend the linear minimum mean-square error (LMMSE) detector to handle the improper Gaussian noise arising from the receiver IQ imbalance. Simulation results demonstrate that the proposed LMMSE detector effectively compensates for the receiver hardware impairments.
\end{abstract}

\section{Introduction}

Next-generation  wireless communication systems are expected to deliver reliable performance in challenging environments such as high-mobility scenarios. Conventional multicarrier techniques such as orthogonal frequency division multiplexing (OFDM) are effective against multipath propagation but face inherent limitations under time-varying channels. These challenges highlight the need for new waveforms that can effectively handle both delay and Doppler shifts introduced by channel dispersiveness and high mobility. 

Affine frequency division multiplexing (AFDM) is a novel waveform that has recently gained attention as a promising waveform to overcome the limitations of conventional multicarrier systems \cite{bemani2023afdm}. 
In the introductory work on AFDM, Bemani \textit{et~al.}~\cite{bemani2023afdm} proposed a maximum ratio combining (MRC)-based iterative decision feedback equalizer (DFE) that benefits from the sparse structure of the AFDM channel matrix to reduce the detection complexity. Additionally, Wu \textit{et~al.}~\cite{wu2024afdmmp} exploited the AFDM sparse channel matrix and proposed a detection algorithm based on message passing that  was shown to be superior to the MRC-based scheme in \cite{bemani2023afdm}. Other detection schemes that exploited the sparsity of the AFDM channel matrix are discussed in \cite{tao2024afdmim, xu2024mbuamp}.

Recently, several studies explored different aspects of AFDM. For example, channel estimation for AFDM has been explored in, \cite{benzine2023afdm, cao2024afdm}, the integration of AFDM with index modulation for multiple antenna has been investigated in \cite{tao2024afdmim}, the applications of AFDM in integrated sensing and communications (ISAC) are explored in \cite{temiz2025afdmplim}, and the evaluation of its ambiguity function with random data symbols is studied in \cite{bedeer2025AF}.
While the aforementioned studies \cite{bemani2023afdm, wu2024afdmmp, tao2024afdmim, xu2024mbuamp, benzine2023afdm, cao2024afdm, temiz2025afdmplim, bedeer2025AF} highlight the potential of AFDM as a promising waveform, they all assumed ideal radio frequency (RF) hardware. Investigating the effects of hardware impairments is important as their mitigation is critical for the efficient operation of communication systems in practical scenarios. 

In this paper, we analyze the impact of receiver hardware impairments on AFDM signals. In particular, we consider the receiver in-phase and quadrature  (IQ) imbalance and residual carrier frequency offset (CFO) error. For brevity, we refer to the receiver IQ imbalance simply as IQ imbalance in the remainder of the paper. 
Our analysis shows that the receiver IQ imbalance may not preserve the sparsity of the AFDM effective channel matrix because of the complex-conjugate operator of the discrete affine Fourier transform (DAFT). Moreover,  residual CFO errors cause energy leakage in the effective channel matrix in the affine domain. 
To address these hardware impairments, we propose a modified  linear minimum mean-square error (LMMSE) detector that compensates for the improper Gaussian noise associated with IQ imbalance. 
Simulation results demonstrate that the proposed LMMSE detector effectively compensates for the receiver hardware impairments.

The remainder of this paper is organized as follows. In Section~II, we present the AFDM system model in the presence of residual CFO error and IQ imbalance. In Section~III, we analyze the effect of the IQ imbalance and residual CFO error on the sparsity of the AFDM channel matrix. Section~\ref{sec:detector} presents the LMMSE detector. Simulation results are provided in Section~V, and conclusions are drawn in Section~VI.

\textbf{Notation:} Scalars are denoted by lowercase letters (e.g., $x$), vectors by bold lowercase letters (e.g., $\mathbf{x}$), and matrices by bold uppercase letters (e.g., $\mathbf{X}$). The operators $(\cdot)^*$, $(\cdot)^{\mathsf{T}}$, and $(\cdot)^{\mathsf{H}}$ denote the complex conjugate, transpose, and Hermitian (conjugate transpose), respectively. The real and imaginary parts of complex numbers are denoted by $\Re\{\cdot\}$ and $\Im\{\cdot\}$. 
A univariate complex Gaussian random variable with mean $\mu$ and variance $\sigma^2$ is denoted by $\mathcal{CN}(\mu,\sigma^2)$. A multi-variate complex Gaussian random vector with mean $\boldsymbol{\mu}$, covariance $\mathbf{C}$, and pseudo-covariance $\mathbf{P}$ is denoted by $\mathcal{CN}(\boldsymbol{\mu},\mathbf{C},\mathbf{P})$. The set of complex numbers is represented by $\mathbb{C}$, with $\mathbb{C}^{N}$ denotes the set of $N$-dimensional complex vectors, and $\mathbb{C}^{N\times M}$ denotes the set of $N\times M$ complex matrices. We use $\mathbf{I}_N$ to denote the $N \times N$ identity matrix. The modular operation is denoted by $(a)_N$. The operator $\operatorname{diag}(\cdot)$ denotes a diagonal matrix whose diagonal entries are given by the argument, and the operator $\operatorname{tr}(\cdot)$ denotes the trace of a square matrix.
The $(m,\ell)$-th element of a matrix $\mathbf{M}$ is denoted by $[\mathbf{M}]_{m,\ell}$.

\section{System Model}
\label{sec:system_model}

We consider an AFDM system 
where the information bits sequence is mapped to blocks of $N$ complex data symbols $\mathbf{x} = [x[0], \cdots, x[N-1]]^{\mathsf{T}} \in \mathbb{C}^{N \times 1}$, drawn from a proper $M$-ary QAM constellation (i.e., $\E(x[n]) = 0$ and $\E(x[n]^2) = 0$) with normalized unit power (i.e., $\E(|x[n]|^2) = 1$). These blocks  of  $N$ data symbols are then modulated onto $N$ AFDM chirp subcarriers, and the discrete-time domain signal $\mathbf{s} = [s[0], \cdots, s[N-1]]^{\mathsf{T}} \in \mathbb{C}^{N \times 1}$ is obtained by applying the inverse discrete affine Fourier transform (IDAFT) as follows~\cite{bemani2023afdm}
\begin{equation}
s[n] = \sum_{m=0}^{N-1} x[m] \, \phi_n[m], \quad n = 0, \ldots, N-1,
\label{eq:tx_signal}
\end{equation}
where $\phi_n[m]$ denotes the AFDM kernel function given by
\begin{equation}
\phi_n[m] = \frac{1}{\sqrt{N}} \, e^{j 2 \pi \left( c_1 n^2 + c_2 m^2 + \frac{n m}{N} \right)} ,
\label{eq:phi_kernel}
\end{equation}
and $c_1$ and $c_2$ are the AFDM design parameters~\cite{bemani2023afdm}. The signal in \eqref{eq:tx_signal} can also be expressed compactly in a matrix form as~\cite{bemani2023afdm}
\begin{equation}
    \mathbf{s} = \mathbf{A}^{\mathsf{H}}\mathbf{x},
    \label{eq:tx_matrix}
\end{equation}
where 
$\mathbf{A}$ is the DAFT matrix given by $\mathbf{A} = \boldsymbol{\Lambda}_{c_2} \mathbf{F} \boldsymbol{\Lambda}_{c_1}$, $\mathbf{F}$ is the  unitary DFT matrix, 
and $\boldsymbol{\Lambda}_c = \operatorname{diag}\!\left( e^{-j 2\pi c n^2}\right), \; n=0,\ldots,N-1,$
where $c = \{c_1, c_2\}$. To mitigate the effects of multipath propagation, a chirp-periodic prefix (CPP) of length at least equal to the number of channel taps is added to the signal before transmission as described in \cite{bemani2023afdm}.
The signal is then transformed to the continuous-time domain using digital-to-analog (DAC), and transmitted using an RF front-end  assumed ideal, and hence, does not introduce any hardware impairments.  
We consider a multipath channel with $P$ resolvable paths, each characterized by a complex channel coefficient $h_p$, delay $\tilde{\tau}_p$ (measured in seconds), and Doppler shift $\tilde{f}_p$ (measured in Hertz). The channel impulse response is given by
\begin{equation}
h(t,\tau) = \sum_{p=1}^{P} h_p \, \delta(\tau - \tilde{\tau}_p) \, e^{-j 2 \pi \tilde{f}_p t}.
\label{eq:ch_impulse}
\end{equation}
The receiver RF front-end is assumed to be non-ideal. In particular, it suffers from CFO with respect to the transmitter carrier frequency $f_c$, and also suffers from amplitude and phase mismatches between the I and Q branches. CFO is deterministic and results in a linear phase rotation across all chirps. In this paper, we do not discuss the CFO estimation due to space limitations, but we assume that the CFO estimation is not perfect and there will be residual CFO errors. 
The IQ imbalance is characterized by the complex scalar parameters $\mu$ and $\nu$ defined as: $\mu = \cos(\phi) + j\psi \sin(\phi)$ and $\nu = \psi \cos(\phi) - j \sin(\phi)$,
where $\phi$ and $\psi$ are the phase orthogonality mismatch and the amplitude mismatch between the I and Q branches ~\cite{tandur2007joint}. 
The effect of the IQ imbalance at the receiver results in the superposition of the received signal scaled by $\mu$ and its complex-conjugate scaled by $\nu$. That being said,  the equivalent baseband received signal after sampling with $T_s = 1/B$, where $B$ is the AFDM signal bandwidth, and after removing the CPP, is written as
\begin{equation}
	\label{eq:cfoiq_td_wrap}
	\begin{aligned}
		r_{\text{CFO+IQ}}[n]
		&= \mu\, e^{-j2\pi \varepsilon n}\, \sum_{p=1}^{P} h_p\, s[n-\tau_p]\,
		e^{-j2\pi f_p n}\, \\
		&\quad + \nu\, e^{+j2\pi \varepsilon n}\, \sum_{p=1}^{P} h_p^{*}\, s^{*}[n-\tau_p]\,
		e^{+j2\pi f_p n}\, \\
		&\quad + \mu\, e^{-j2\pi \varepsilon n}\, w[n]\,
		+ \nu\, e^{+j2\pi \varepsilon n}\, w^{*}[n]\,,
	\end{aligned}
\end{equation}
where $\tau_p = \tilde{\tau}_p/T_s$, $f_p = \tilde{f}_p T_s$, and  $w[n] \sim \mathcal{CN}(0,\sigma_n^2)$ is a proper additive white Gaussian noise with zero-mean and variance $\sigma_n^2$.  The normalized residual CFO error is defined as $\varepsilon = \widehat{f}_{c,\Delta} T_s - f_{c,\Delta} T_s$
where $f_{c,\Delta}$ and $\widehat{f}_{c,\Delta}$ are the true CFO and estimated CFO, respectively.
While the statistics of the residual CFO errors $\varepsilon$ in general depend on the CFO estimator, which is not discussed due to space limitation, we assume in this paper that the residual CFO error $\varepsilon$ follows a Gaussian distribution $\varepsilon \sim \mathcal{N}(0, \sigma_{\varepsilon}^2)$ as in \cite{almradi2015spectral}. The discrete-time received signal in the presence of residual CFO error and IQ imbalance in \eqref{eq:cfoiq_td_wrap} is re-expressed as
\begin{equation}
\label{eq:cfoiq_mx_wrap}
\begin{aligned}
\mathbf{r}_{\text{CFO+IQ}}
&= \mu\,\boldsymbol{\Delta}_{\varepsilon}
   \Bigg(\sum_{p=1}^{P} h_p\, \boldsymbol{\Gamma}_{\text{CPP}_p}\,
   \boldsymbol{\Delta}_{f_p}\,\boldsymbol{\Pi}^{\,\tau_p}\,\mathbf{s}
   + \mathbf{w}\Bigg) \\
&\quad + \nu\,\boldsymbol{\Delta}_{-\varepsilon}
   \Bigg(\sum_{p=1}^{P} h_p^{*}\,\boldsymbol{\Gamma}_{\text{CPP}_p}^{*}\,
   \boldsymbol{\Delta}_{-f_p}\,\boldsymbol{\Pi}^{\,\tau_p}\,\mathbf{s}^{*}
   + \mathbf{w}^{*}\Bigg),
\end{aligned}
\end{equation}
where $\mathbf{w} \sim \mathcal{CN}(\mathbf{0}, \sigma^2_n \mathbf{I}_N)$, $\boldsymbol{\Delta}_{\lambda} = \operatorname{diag}\!\left(e^{-j2\pi \lambda n}\right)$ and $\lambda = \{\pm \varepsilon, \pm f_p \}$ for $n = 0, \ldots, N-1$, $\boldsymbol{\Gamma}_{\text{CPP}_p}$ reduces to  $\mathbf{I}_N$ when $2 N c_1$ is an integer  \cite{bemani2023afdm}, which we follow in the paper,
and $\boldsymbol{\Pi}$ denotes the forward cyclic-shift matrix defined as 
\begin{equation}
\label{eq:Pi}
\boldsymbol{\Pi} = 
\begin{bmatrix}
0 & \cdots & 0 & 1 \\
1 & \cdots & 0 & 0 \\
\vdots & \ddots & \ddots & \vdots \\
0 & \cdots & 1 & 0
\end{bmatrix}_{N \times N}.
\end{equation}
To detect the transmit $M$-ary QAM symbols in the affine domain, we apply the DAFT on $\mathbf{r}_{\text{CFO+IQ}}$, i.e., $\mathbf{A}\,\mathbf{r}_{\text{CFO+IQ}}$,  which results in
\begin{align}
\mathbf{y}
&= \mu\,\mathbf{A}\Bigg(\boldsymbol{\Delta}_{\varepsilon}
      \sum_{p=1}^{P} h_p\,
      \boldsymbol{\Gamma}_{\text{CPP}_p}\,
      \boldsymbol{\Delta}_{f_p}\,
      \boldsymbol{\Pi}^{\,\tau_p}\,\mathbf{A}^{\mathsf{H}}\,\mathbf{x}\Bigg) \nonumber\\
&\quad + \nu\,\mathbf{A}\,\Bigg(\boldsymbol{\Delta}_{\varepsilon}
      \sum_{p=1}^{P} h_p\,
      \boldsymbol{\Gamma}_{\text{CPP}_p}\,
      \boldsymbol{\Delta}_{f_p}\,
      \boldsymbol{\Pi}^{\,\tau_p}\,\mathbf{A}^{\mathsf{H}}\;\mathbf{x}\Bigg)^{*}
      \;+\; \mathbf{w}',
\label{eq:daft_iqcfo}
\end{align}
where the zero-mean improper noise $\mathbf{w}'$ is given as
\begin{align}
\mathbf{w}'
&= \mu\,\mathbf{A}\,\boldsymbol{\Delta}_{\varepsilon}\,\mathbf{w}
   + \nu\,\mathbf{A}\,\boldsymbol{\Delta}_{-\varepsilon}\,\mathbf{w}^{*},
\label{eq:w_dash}
\end{align}
with the covariance $\mathbf{C}_{\mathbf{w}'}=\E({\mathbf{w}'}{\mathbf{w}'}^\mathsf{H}) = (|\mu|^2+\!|\nu|^2) \sigma_n^2 \mathbf{I}_N$ and pseudo-covariance $\mathbf{P}_{\mathbf{w}'} = \E({\mathbf{w}'}{\mathbf{w}'}^\mathsf{T}) = 2\mu\nu\,\sigma_n^2\mathbf{A}\mathbf{A}^{\mathsf{T}}$. In other words, the improper noise vector $\mathbf{w}'$ is distributed as $\mathcal{CN} \left(\mathbf{0}, (|\mu|^2+\!|\nu|^2) \sigma_n^2 \mathbf{I}_N,2\mu\nu\,\sigma_n^2\mathbf{A}\mathbf{A}^{\mathsf{T}} \right)$. Then we can use a detector that detects the transmit $M$-ary QAM symbols $\mathbf{x}$ and hence the information bits from $\mathbf{y}$, in the presence of residual CFO error, IQ imbalance, and improper Gaussian noise.

\section{Analysis of Residual CFO Error and IQ Imbalance on AFDM Signals}
In this section, we analyze the effects of the residual CFO error and IQ imbalance on the AFDM signal and its effective channel matrix. Since the IQ imbalance results in the superposition of a scaled version of the received signal and a scaled version of its complex-conjugate, we first introduce the complex-conjugate property of the DAFT as follows.

\textit{The complex-conjugate property of the DAFT:} For the DAFT defined as $\mathbf{x} = \mathbf{A}\mathbf{s}$, the DAFT of the complex-conjugate of the sequence $\mathbf{s}$, i.e., $\textup{DAFT}(\mathbf{s}^{*})$, equals applying the operator $\mathbf{A}\mathbf{A}^{\mathsf{T}}$ to the complex-conjugate of the DAFT of the original sequence $\mathbf{s}$, i.e., $\left(\textup{DAFT}(\mathbf{s})\right)^*$.

\textit{Proof:} The DAFT of the time domain vector $\mathbf{s}$ is defined as $\mathbf{x} = \mathbf{A}\mathbf{s}$. Similarly, the DAFT of the time-domain vector $\mathbf{s}^*$ is defined as $\breve{\mathbf{x}} = \mathbf{A}\mathbf{s}^* = \mathbf{A} (\mathbf{A}^{\mathsf{H}}\mathbf{x})^* = \mathbf{A}\mathbf{A}^{\mathsf{T}}\mathbf{x}^*$. \hfill $\blacksquare$

\begin{figure}[t]
  \centering
  \begin{subfigure}[t]{0.48\linewidth}
    \includegraphics[width=\linewidth]{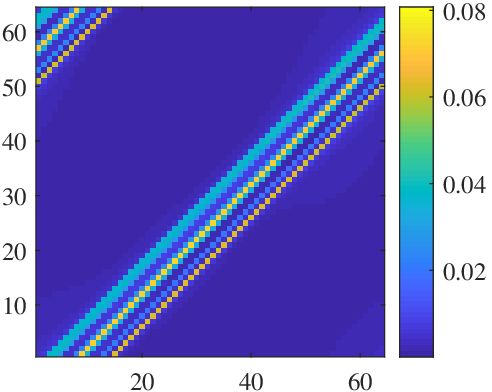}
    \subcaption{$\sigma^2_{\varepsilon}=10^{-3}$}\label{fig:heffA}
  \end{subfigure}\hfill
  \begin{subfigure}[t]{0.48\linewidth}
    \includegraphics[width=\linewidth]{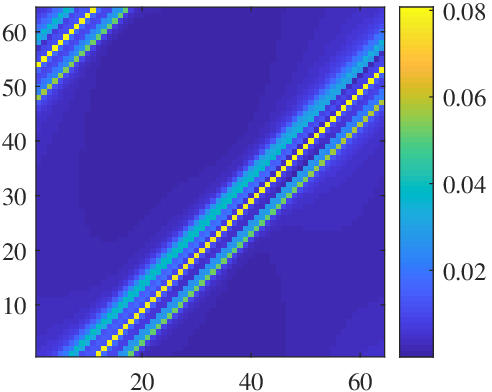}
    \subcaption{$\sigma^2_{\varepsilon}=10^{-1}$}\label{fig:heffB}
  \end{subfigure}
  \caption{Comparison of $\mathbf{H}_{\mathrm{eff}}$ for two residual CFO error variance values, for $M$ = 4, \(N=64\), \(P=3\), \(2Nc_{1}=5\), and \(c_{2}=0\).}
  \label{fig:heff-sigma-compare}
\end{figure}

The signal $\mathbf{y}$ in the affine domain in \eqref{eq:daft_iqcfo} can be simplified with the help of the complex-conjugate property of the DAFT as
\begin{IEEEeqnarray}{RCL}
	\mathbf{y}&{}={} \mu \mathbf{H}_{\mathrm{eff}} \mathbf{x} + \nu\,\mathbf{A}\mathbf{A}^{\mathsf{T}}\mathbf{H}_{\mathrm{eff}}^{*}\,\mathbf{x}^{*} + \mathbf{w}', \label{eq:in_out}
\end{IEEEeqnarray}
where 
\begin{align}
\mathbf{H}_{\mathrm{eff}}
&= \mathbf{A}\,\boldsymbol{\Delta}_{\varepsilon}\,\left(\sum_{p=1}^{P} h_p\,
  \boldsymbol{\Gamma}_{\text{CPP}_p}\,
  \boldsymbol{\Delta}_{f_p}\,
  \boldsymbol{\Pi}^{\,\tau_p}\right)\,\mathbf{A}^{\mathsf{H}}.
\label{eq:H_eff-def}
\end{align}
Please note that the effective channel matrix $\mathbf{H}_{\mathrm{eff}}$ as defined in \eqref{eq:H_eff-def} is different from its  definition in \cite{bemani2023afdm} as it now includes the effect of the residual CFO errors $\boldsymbol{\Delta}_{\varepsilon}$. As illustrated in Fig.~\ref{fig:heff-sigma-compare}, the impact of residual CFO error on $\mathbf{H}_{\mathrm{eff}}$ is mainly an energy leakage that depends on $\sigma_{\varepsilon}^2$. As the residual CFO error variance $\sigma_{\varepsilon}^2$ increases, the energy leakage increases and may eventually affect the path resolvability in the affine domain.

To study the effect of the  IQ imbalance on the received AFDM signal, we consider the case of no CFO, i.e., $\varepsilon=0$. As can be seen in \eqref{eq:in_out}, the effects of the IQ imbalance on the AFDM system model are: (i) scaling of $\mathbf{H}_{\mathrm{eff}} \mathbf{x}$ by $\mu$, which is equivalent to attenuation and phase rotation of $\mathbf{H}_{\mathrm{eff}} $; (ii) scaling, by $\nu$, and applying the complex-conjugate operator of the DAFT, i.e., $\mathbf{A}\mathbf{A}^{\mathsf{T}}$, to $\left(\mathbf{H}_{\mathrm{eff}} \mathbf{x}\right)^*$, which is equivalent to applying the operator $\mathbf{A}\mathbf{A}^{\mathsf{T}}$ to an attenuation and phase-rotated $\mathbf{H}^*_{\mathrm{eff}}$; and (iii) having improper complex-Gaussian noise $\mathbf{w}'$. 

In the following, we discuss the effect of the complex-conjugate operator of the DAFT, i.e., $\mathbf{A}\mathbf{A}^{\mathsf{T}}$, on the sparsity of the AFDM effective channel matrix. The operator $\mathbf{A}\mathbf{A}^{\mathsf{T}}$ can be expressed as
\begin{IEEEeqnarray}{RCL}
	\mathbf{A}\mathbf{A}^{\mathsf{T}}&{}={}& \boldsymbol{\Lambda}_{c_2} \mathbf{F} \boldsymbol{\Lambda}_{2 c_1} \mathbf{F}  \boldsymbol{\Lambda}_{c_2} \label{eq:operator}
\end{IEEEeqnarray}
by noting that $\boldsymbol{\Lambda}_{c} = \boldsymbol{\Lambda}_{c}^{\mathsf{T}}$ for phase-diagonal matrices, $\mathbf{F} = \mathbf{F}^{\mathsf{T}}$ for the unitary DFT matrix, and $\boldsymbol{\Lambda}_{2 c_1} = \boldsymbol{\Lambda}_{c_1}^2$. 
Note that when $c_1 = c_2 = 0$, and hence, the DAFT reduces to the DFT and the AFDM reduces to OFDM, the operator $\mathbf{A}\mathbf{A}^{\mathsf{T}} = \mathbf{F}^2$ reduces to the following permutation matrix: 
\begin{equation}
\begin{aligned}
[\mathbf{T}]_{m, \ell} \;&=\;
\begin{cases}
1, & \text{if } (m+\ell)_N=0, \\[0.5ex]
0, & \text{otherwise}.
\end{cases}
\end{aligned}
\end{equation}
Such permutation matrix $\mathbf{T}$ in case of the DFT or OFDM results in the known effect of coupling the OFDM subcarrier and their mirrored counterparts due to the IQ imbalance. For the DAFT or AFDM, the IQ imbalance will not result in simple mirroring of chirps and their counterparts, and this is due to the possible no-sparsity of the operator $\mathbf{A}\mathbf{A}^{\mathsf{T}}$. To explain this issue, we set $c_2=0$ \cite{bemani2023afdm}, then the operator $\mathbf{A}\mathbf{A}^{\mathsf{T}}$ is simplified as $	\mathbf{A}\mathbf{A}^{\mathsf{T}} = \mathbf{F} \boldsymbol{\Lambda}_{2 c_1} \mathbf{F}$, 
and the element $(m,\ell)$ of complex-conjugate operator $\mathbf{A}\mathbf{A}^{\mathsf{T}}$ is written as 
\begin{align}
[\mathbf{F}\mathbf{\Lambda}_{2c_1}\mathbf{F}]_{m,\ell}
&=\sum_{n=0}^{N-1}[\mathbf{F}]_{m,n}\,[\mathbf{\Lambda}_{2c_1}]_{n,n}\,[\mathbf{F}]_{n,\ell} \nonumber\\[0.25em]
&=\frac{1}{N}\sum_{n=0}^{N-1}
\exp\!\left(-j\frac{2\pi}{N}\big((m+\ell)\,n+2Nc_1n^2\big)\right).
\label{eq:FLF}
\end{align}

\begin{figure*}[!t]
  \centering
  \subfloat[$2Nc_1=5$, left: $\mathbf{A}\mathbf{A}^{\mathsf{T}}$, and right: $\mathbf{A}\mathbf{A}^{\mathsf{T}} \mathbf{H}_{\mathrm{eff}}^{*}$]{%
    \includegraphics[width=0.49\textwidth]{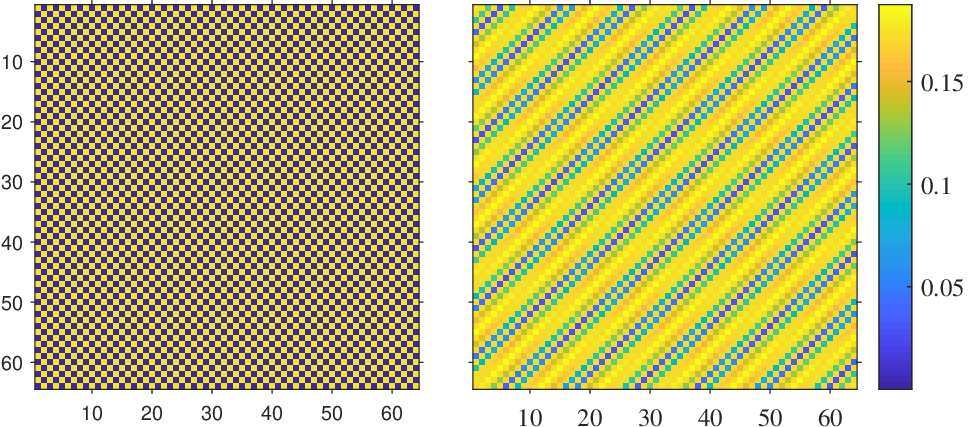}%
    \label{fig:2Nc1-odd}}
  \hfill
  \subfloat[$2Nc_1=10$, left: $\mathbf{A}\mathbf{A}^{\mathsf{T}}$, right: $\mathbf{A}\mathbf{A}^{\mathsf{T}} \mathbf{H}_{\mathrm{eff}}^{*}$]{%
    \includegraphics[width=0.49\textwidth]{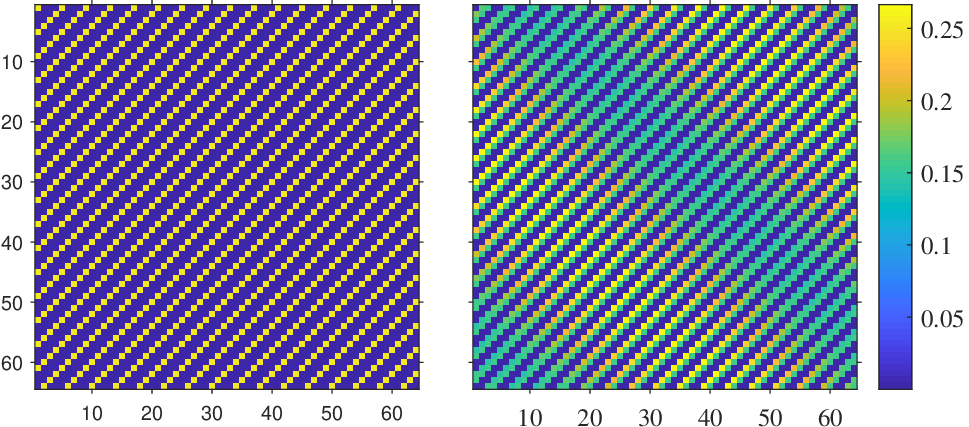}%
    \label{fig:2Nc1-odd}}
  \caption{The operator $\mathbf{A}\mathbf{A}^{\mathsf{T}}$ and its effect on $\mathbf{H}_{\mathrm{eff}}^{*}$ for odd and even integer values of $2Nc_1$, for $M$ = 4,  \(N=64\), \(P=3\), \(c_{2}=0\),
  \(\phi=8^{\circ}\), \(\psi=0.1\), \(\sigma_{\varepsilon}^2=0\).}
  \label{fig:complex-conjugate-operator}
\end{figure*}

To investigate if  $\mathbf{A}\mathbf{A}^{\mathsf{T}}$ is sparse or dense, we need to evalute the sum in \eqref{eq:FLF} for each element $(m,\ell)$, $m,\ell\in\{0,1,\dots,N-1\}$, and identify the conditions under which these elements are zero or non-zero. One can observe that for each element $(m,\ell)$, we need to add $N$ complex exponentials with fixed magnitude, so one can also re-express \eqref{eq:FLF} as
\begin{IEEEeqnarray}{RCL}\label{eq:sum_2}
[\mathbf{F}\mathbf{\Lambda}_{2c_1}\mathbf{F}]_{m,\ell} &{}={}& \frac{1}{N} \sum_{n=0}^{N/2-1} S_n + S_{n+N/2},
\end{IEEEeqnarray}
where $S_n = \exp\!\left(-j\frac{2\pi}{N}\big((m+\ell)\,n+2Nc_1n^2\big)\right)$. Then by exploring the relation between $S_n$ and $S_{n+N/2}$, one can tell if $[\mathbf{F}\mathbf{\Lambda}_{2c_1}\mathbf{F}]_{m,\ell}$ is zero or not. That being said, for an even value of $N$ {and an integer value of $2Nc_1$}, one can show that
\begin{align}
\frac{S_{n+N/2}}{S_n} 
&= \frac{\exp{\left(-j\frac{2\pi}{N}(m+\ell)(n+\frac{N}{2}) + (2Nc_1)(n+\frac{N}{2})^2\right)}}{\exp{\left(-j\frac{2\pi}{N}(m+\ell)n + (2Nc_1)n^2\right)}},\notag\\
&= (-1)^{m+\ell}.
\label{eq:S_ratio}
\end{align}
Hence, it is clear that if $m + \ell$ is odd, then  $S_n = - S_{n+N/2}$, and $[\mathbf{F}\mathbf{\Lambda}_{2c_1}\mathbf{F}]_{m,\ell} = 0$. On the other hand, if $m + \ell$ is even, then  $a_n = a_{n+N/2}$, and $[\mathbf{F}\mathbf{\Lambda}_{2c_1}\mathbf{F}]_{m,\ell}$ may or may not equal to zero. 

When  $m + \ell$ is even, and hence, $a_n = a_{n+N/2}$, the term $m + \ell = 2 q$, where $q$ can be odd or even, and the sum in \eqref{eq:sum_2} is rewritten as
\begin{IEEEeqnarray}{RCL}\label{eq:sum_3}
[\mathbf{F}\mathbf{\Lambda}_{2c_1}\mathbf{F}]_{m,\ell} &{}={}& \frac{1}{N} \sum_{n=0}^{N/2-1} 2 S_n =  \frac{1}{N/2} \sum_{n=0}^{N/2-1}  S_n , \nonumber \\
&{}={}& \frac{1}{N/2} \sum_{n=0}^{N/2-1} e^{-j\frac{2\pi}{N/2}\big( q\,n+Nc_1n^2\big)}.
\end{IEEEeqnarray}
Note that the quadratic phase term has $N c_1$ instead of $2 N c_1$. This sum in \eqref{eq:sum_3} can be treated similar to  \eqref{eq:FLF} and \eqref{eq:sum_2}, and we can show that $[\mathbf{F}\mathbf{\Lambda}_{2c_1}\mathbf{F}]_{m,\ell} = 0$ when $q$ is odd and $N c_1$ is an integer; otherwise, it may or may not be zero. 

When $q$ in \eqref{eq:sum_3} is even, and similar to the previous discussion, we can replace it with $q = 2 q'$, where $q'$ can be odd or even. By noting that the quadratic phase term is now $N c_1/2$ instead of $N c_1$. Then we can proceed to show that $[\mathbf{F}\mathbf{\Lambda}_{2c_1}\mathbf{F}]_{m,\ell} = 0$ when $q'$ is odd and $N c_1/2$ is integer; otherwise, it may or may not be zero. By induction, one can conclude that $[\mathbf{F}\mathbf{\Lambda}_{2c_1}\mathbf{F}]_{m,\ell} = 0$ if and only if $m + \ell$ is even and $2 N c_1$ is divisible by a higher power of 2 than $m+\ell$, i.e., $2 N c_1$ has a higher 2-adic valuation than that of the even $m + \ell$.

For the case, when $2 N c_1$ has the same 2-adic valuation as $m + \ell$, i.e., $2 N c_1 = 2^{a}v$ and $m + \ell = 2^{a}u$ where $v$ and $u$ are odd numbers, then the sum in \eqref{eq:FLF} can be evaluated as
\begin{align}
[\mathbf{F}\mathbf{\Lambda}_{2c_1}\mathbf{F}]_{m,\ell}
&= \frac{1}{N}\sum_{n=0}^{N-1}
\exp\!\left(-j\frac{2\pi}{N 2^{-a}}\big(u\,n + v\,n^{2}\big)\right),
\label{eq:FLF_even}
\end{align}
which equals zero, following the treatment of \eqref{eq:FLF} and \eqref{eq:sum_2}, as $u$ is odd and $v$ is integer. 

To conclude this discussion, the complex-conjugate operator of the DAFT, i.e., $\mathbf{A}\mathbf{A}^{\mathsf{T}}$, has zero entries if either of the following conditions is satisfied: 
\begin{enumerate}[label=(\roman*)]
	\item $m + \ell$ is odd regardless of the integer value of $2N c_1$,
	\item $m + \ell$ is even and {the 2-adic valuation of $2Nc_{1}$ is greater than or equal to that of $m + \ell$}.
\end{enumerate} 
If the two conditions are not satisfied, then the entries of the operator $\mathbf{A}\mathbf{A}^{\mathsf{T}}$ are non-zeros. This means that in general the operator $\mathbf{A}\mathbf{A}^{\mathsf{T}}$ may be a dense matrix. Figure 
  \ref{fig:complex-conjugate-operator} depicts this important observation. Note that since  the operator $\mathbf{A}\mathbf{A}^{\mathsf{T}}$ does not have a permutation structure (as in DFT or OFDM) or a guaranteed sparse structure, this  has the potential of destroying the sparsity of the AFDM channel matrix and spread its energy; hence, making the tailored detectors of AFDM that relies on its sparsity, e.g., MRC and message passing, unfeasible from a complexity perspective. However, for non-severe IQ imbalance, i.e., $|\nu| \ll |\mu|$, the sparsity of the AFDM channel matrix may be maintained.

\section{Proposed LMMSE Detector}\label{sec:detector}
In this section, we extend the LMMSE detector to accomodate the improper Gaussian noise in \eqref{eq:w_dash}, i.e., $\mathbf{w}' \sim \mathcal{CN} \left(\mathbf{0}, (|\mu|^2+\!|\nu|^2) \sigma_n^2 \mathbf{I}_N,2\mu\nu\,\sigma_n^2\mathbf{A}\mathbf{A}^{\mathsf{T}} \right)$, associated with the IQ imbalance.

The improper noise vector $\mathbf{w}'$ has correlated real and imaginary parts with unequal variances as shown in  the pseudo-covariance $\mathbf{P}_{\mathbf{w}'} = 2\mu\nu\,\sigma_n^2\mathbf{A}\mathbf{A}^{\mathsf{T}}$. Hence, any developed detectors of AFDM in the presence of IQ imbalance must be re-designed to accommodate the improper noise; otherwise, they are sub-optimal.

One possible approach to design detectors that accommodate the improper Gaussian noise is widely linear processing ~\cite{huemer2013widely} that jointly processes the received signal and its conjugate, i.e., $\mathbf{y}$ and $\mathbf{y}^*$. In this paper, we resort to an equivalent approach that processes the real and imaginary parts of the received signal $\mathbf{y}$ as described below. The real and imaginary parts $\mathbf{y}_{\mathrm{R}}$ and $\mathbf{y}_{\mathrm{I}}$, respectively, of $\mathbf{y}$ in \eqref{eq:in_out} can be written as
\begin{align}
\mathbf{y}_{\mathrm{R}} 
&= \Re\!\bigl\{\mu \mathbf{H}_{\mathrm{eff}} 
   + \nu \mathbf{A}\mathbf{A}^{\mathsf{T}}\,\mathbf{H}_{\mathrm{eff}}^{*}\bigr\}\,
   \mathbf{x}_{\mathrm{R}} \nonumber\\
&\quad - \Im\!\bigl\{\mu \mathbf{H}_{\mathrm{eff}} 
   - \nu \mathbf{A}\mathbf{A}^{\mathsf{T}}\,\mathbf{H}_{\mathrm{eff}}^{*}\bigr\}\,
   \mathbf{x}_{\mathrm{I}} +  \mathbf{w}_{\mathrm{R}}'.
\label{eq:y_real}
\end{align}
\begin{align}
\mathbf{y}_{\mathrm{I}}
&= \Im\!\bigl\{\mu \mathbf{H}_{\mathrm{eff}} 
   + \nu \mathbf{A}\mathbf{A}^{\mathsf{T}}\,\mathbf{H}_{\mathrm{eff}}^{*}\bigr\}\,
   \mathbf{x}_{\mathrm{R}} \nonumber\\
&\quad + \Re\!\bigl\{\mu \mathbf{H}_{\mathrm{eff}} 
   - \nu \mathbf{A}\mathbf{A}^{\mathsf{T}}\,\mathbf{H}_{\mathrm{eff}}^{*}\bigr\}\,
   \mathbf{x}_{\mathrm{I}} + \mathbf{w}_{\mathrm{I}}'.
\label{eq:y_imag}
\end{align}
Then, by concatenating the real and imaginary parts of $\mathbf{y}$, $\mathbf{x}$, and $\mathbf{w}'$, one can re-express \eqref{eq:in_out} as
\begin{equation}
\tilde{\mathbf{y}} = \tilde{\mathbf{H}} \tilde{\mathbf{x}} + \tilde{\mathbf{w}},
\end{equation}
where
\begin{equation}
\tilde{\mathbf{H}} =
\small
\begin{bmatrix}
\Re\{\mu \mathbf{H}_{\mathrm{eff}} + \nu \mathbf{A}\mathbf{A}^{\mathsf{T}}\mathbf{H}_{\mathrm{eff}}^{*}\} 
& -\Im\{\mu \mathbf{H}_{\mathrm{eff}} - \nu \mathbf{A}\mathbf{A}^{\mathsf{T}}\mathbf{H}_{\mathrm{eff}}^{*}\} \\[0.5ex]
\Im\{\mu \mathbf{H}_{\mathrm{eff}} + \nu \mathbf{A}\mathbf{A}^{\mathsf{T}}\mathbf{H}_{\mathrm{eff}}^{*}\} 
& \Re\{\mu \mathbf{H}_{\mathrm{eff}} - \nu \mathbf{A}\mathbf{A}^{\mathsf{T}}\mathbf{H}_{\mathrm{eff}}^{*}\}
\end{bmatrix},
\normalsize
\end{equation}
and $\tilde{\mathbf{w}} \sim \mathcal{N} \left(\mathbf{0}, \mathbf{C}_{\tilde{\mathbf{w}}} \right)$ where 
\begin{equation}
\mathbf{C}_{\tilde{\mathbf{w}}}
= \tfrac{1}{2}
\begin{bmatrix}
\Re\{\mathbf{C}_{\mathbf{w}'} + \mathbf{P}_{\mathbf{w}'}\} & -\Im\{\mathbf{C}_{\mathbf{w}'} - \mathbf{P}_{\mathbf{w}'}\} \\[0.5ex]
\Im\{\mathbf{C}_{\mathbf{w}'} + \mathbf{P}_{\mathbf{w}'}\} & \Re\{\mathbf{C}_{\mathbf{w}'} - \mathbf{P}_{\mathbf{w}'}\}
\end{bmatrix}.
\end{equation}
The detector objective now is to estimate $\tilde{\mathbf{x}}$ from $\tilde{\mathbf{y}}$ given the modified effective channel matrix $\tilde{\mathbf{H}}$ in the presence of the noise $\tilde{\mathbf{w}}$.

\begin{figure}[t]
  \centering
  \includegraphics[width=0.9\linewidth]{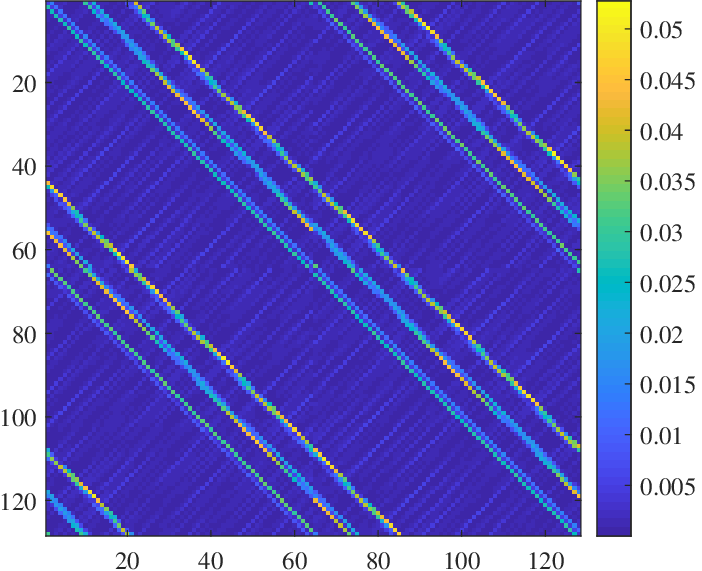}
  \caption{Concatenated effective channel matrix $\tilde{\mathbf{H}}$ for $M$ = 4,  \(N=64\), \(P=3\), \(2Nc_{1}=5\), \(c_{2}=0\),
  \(\phi=8^{\circ}\), \(\psi=0.1\), and \(\sigma_{\varepsilon}^2=10^{-1}\).}
  \label{fig:H-tilde}
\end{figure}
As noted earlier, the effect of IQ imbalance may not maintain the sparsity of the effective channel matrix. This can also be seen when considering the concatenated channel matrix $\tilde{\mathbf{H}}$ where the sparse shifted-diagonal structure is only partially preserved, as illustrated in Fig.~\ref{fig:H-tilde}.  For severe IQ imbalance scenarios, $\tilde{\mathbf{H}}$ will contain a significant number of nonzero elements due to the operator $\mathbf{A}\mathbf{A}^{\mathsf{T}}$. The LMMSE estimator of $\tilde{\mathbf{x}}$, denoted by $\hat{\tilde{\mathbf{x}}}$, can be written as 
\begin{align}
\hat{\tilde{\mathbf{x}}}
&=\bigl(2\mathbf{I}_{2N}
+\tilde{\mathbf{H}}^{\mathsf{T}} \mathbf{C}_{\tilde{\mathbf{w}}}^{-1} \tilde{\mathbf{H}}\bigr)^{-1}
\tilde{\mathbf{H}}^{\mathsf{T}} \mathbf{C}_{\tilde{\mathbf{w}}}^{-1}
\,\tilde{\mathbf{y}}.
\label{eq:tilde-x-estimate-lmmse}
\end{align}
With the help of the Woodbury identity, the LMMSE solution in \eqref{eq:tilde-x-estimate-lmmse} can be simplified as
\begin{align}
\hat{\tilde{\mathbf{x}}}
&=\tfrac{1}{2}\tilde{\mathbf{H}}^{\mathsf{T}} \,\bigl(\tfrac{1}{2}\tilde{\mathbf{H}}\,\tilde{\mathbf{H}}^{\mathsf{T}} 
+\mathbf{C}_{\tilde{\mathbf{w}}}\bigr)^{-1}
\,\tilde{\mathbf{y}}.
\label{eq:tilde-x-estimate-lmmse-short}
\end{align}

\section{Simulation Results}
In this section, we present the bit error rate (BER) results of the proposed LMMSE detector and the low-complexity weighted maximal ratio combining (MRC)-based decision feedback equalizer (DFE) detector introduced in \cite{bemani2023afdm}, in the presence of IQ imbalance and residual CFO error. Please note that the IQ imbalance parameters $\mu$ and $\nu$ are assumed to be known by the receiver during the channel estimation phase. Following the parameters in~\cite{bemani2023afdm}, the complex path gains $h_p$ are modeled as independent complex Gaussian random variables, i.e., $h_p \sim \mathcal{CN}(0, \tfrac{1}{P})$. A carrier frequency of 4\,GHz is also assumed.

\begin{figure}[!t]
    \centering
    \subfloat[Integer Doppler shifts]{%
        \includegraphics[width=\linewidth]{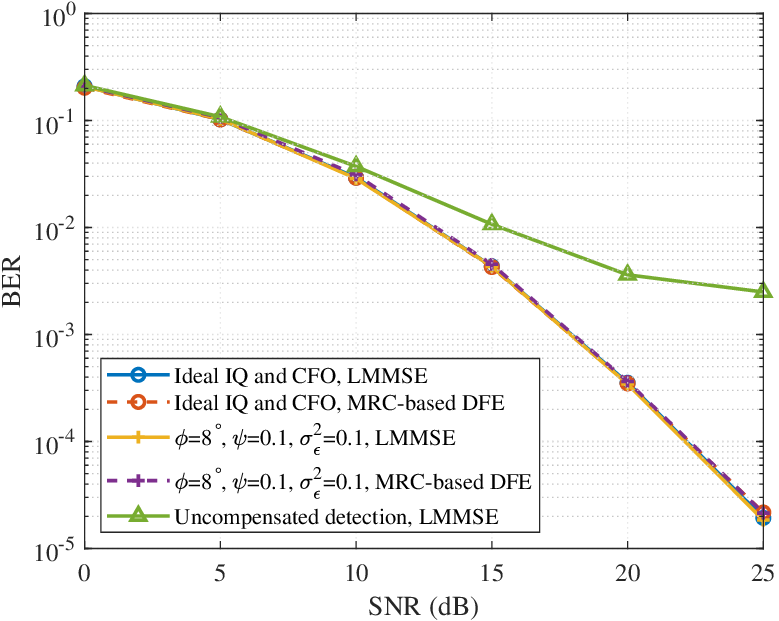}%
        \label{fig:iqcfo-int}}
    \vspace{0.1em}
    \subfloat[Fractional Doppler shifts]{%
        \includegraphics[width=\linewidth]{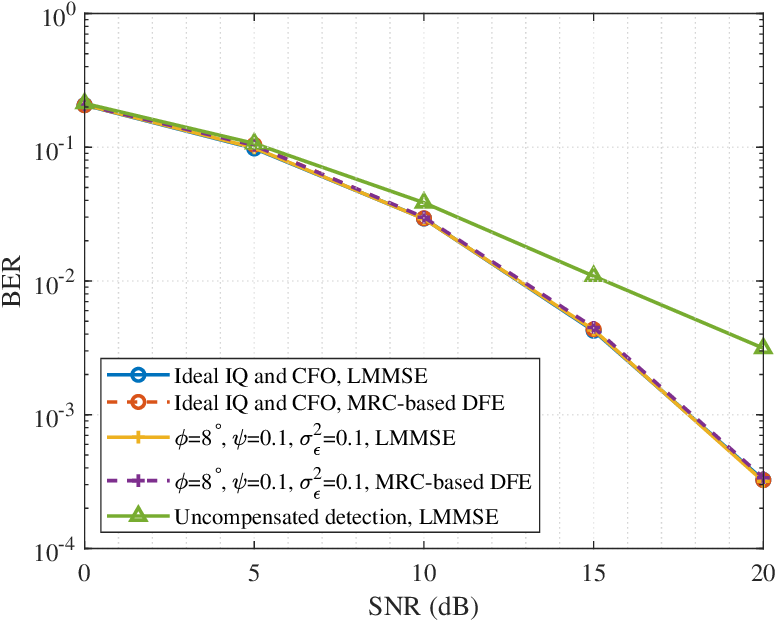}%
        \label{fig:iqcfo-frac}}
    \caption{AFDM performance with IQ imbalance and residual CFO error for $N=256$.}
    \label{fig:iqcfo_256}
\end{figure}

The AFDM system is simulated via Monte Carlo trials using QPSK modulation with $N =$  256 and 128 chirps and $P=3$ delay–Doppler paths. The AFDM parameters are set to $2Nc_{1}=5$ for integer Doppler shifts and $2Nc_{1}=13$ for fractional Doppler shifts, and $c_{2}=0.0001$ for both cases. Following \cite{bemani2023afdm}, the maximum normalized Doppler shift $\alpha_{\max}$ is set to twice the chirp spacing, assuming a maximum speed of 405 km/h, and the path delays were chosen as $\tau_{1}=0$, $\tau_{2}=1$, and $\tau_{3}=2$. The Doppler frequencies were determined based on uniformly distributed angles $\theta_p \in [-\pi,\pi]$, where $p = 1,2,3$. For the integer Doppler case, the Doppler indices $\alpha_p$ were obtained by taking the integer part of $\alpha_{\max}\cos(\theta_p)$, and the corresponding frequencies were given by $f_p = \alpha_p/N$. For the fractional Doppler case, the Doppler frequencies were calculated as $f_p = (\alpha_{\max}\cos(\theta_p))/N$, with the AFDM design parameter $\xi_{\nu}=4$ \cite{bemani2023afdm}.

Fig.~\ref{fig:iqcfo_256} presents the BER performance of the AFDM system for \( N = 256 \) under severe residual CFO error (\( \sigma_{\varepsilon}^2 = 0.1 \)) and severe IQ imbalance (\( \psi = 0.1, \; \phi = 8^{\circ} \)). Both integer and fractional Doppler scenarios are shown. The results indicate that, although IQ imbalance and residual CFO are severe, both detectors effectively mitigate their impact, closely approaching the ideal scenario.

\begin{figure}[!t]
    \centering
    \subfloat[Integer Doppler shifts]{%
        \includegraphics[width=\linewidth]{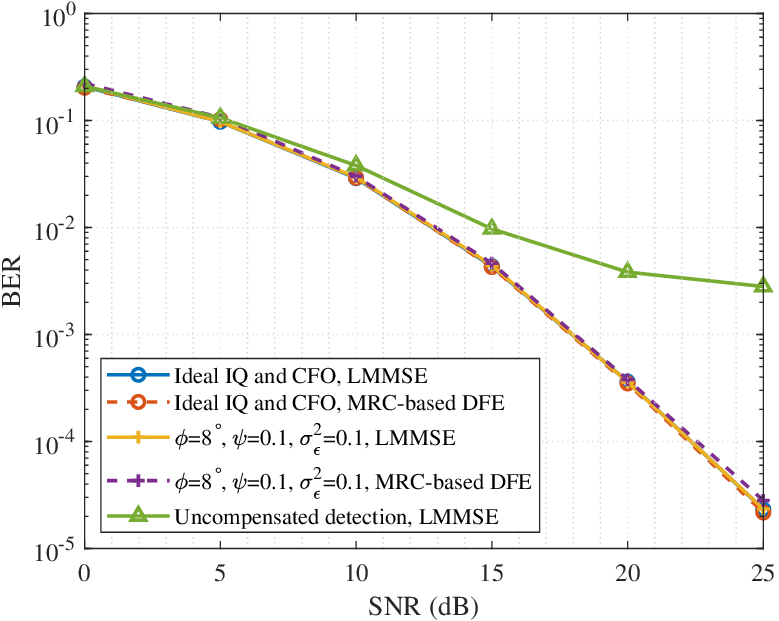}%
        \label{fig:iqcfo-int}}
    \vspace{0.1em}
    \subfloat[Fractional Doppler shifts]{%
        \includegraphics[width=\linewidth]{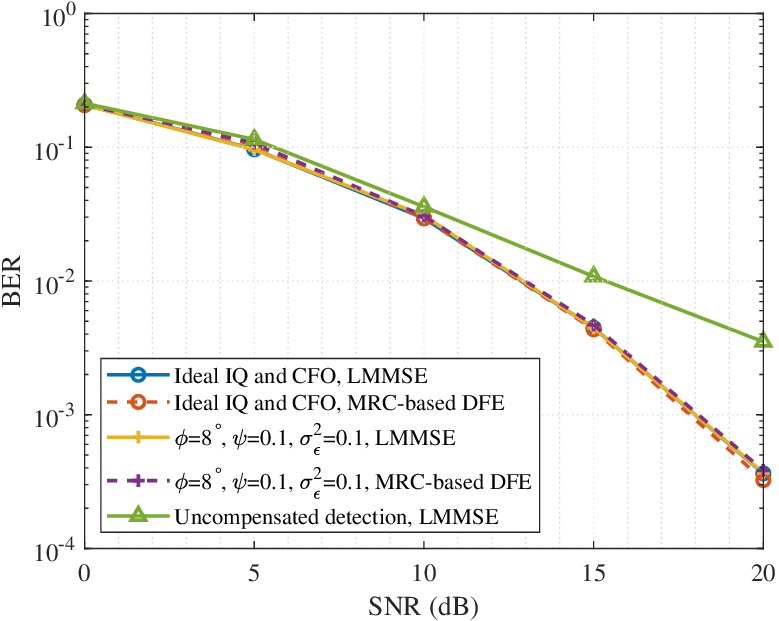}%
        \label{fig:iqcfo-frac}}
    \caption{AFDM performance with IQ imbalance and residual CFO errors for $N=128$.}
    \label{fig:iqcfo_128}
\end{figure}

Fig.~\ref{fig:iqcfo_128} shows the corresponding results for \( N = 128 \) under severe residual CFO error (\( \sigma_{\varepsilon}^2 = 0.1 \)) and severe IQ imbalance (\( \psi = 0.1, \; \phi = 8^{\circ} \)). As with the \( N = 256 \) case, both detectors consistently compensate for residual CFO error and IQ imbalance in both integer and fractional Doppler scenarios, again achieving performance close to the ideal scenario. This demonstrates the robustness of the detectors across different block lengths.

\section{Conclusions}
In this paper, we examined the effect of receiver IQ imbalance and residual CFO error on AFDM signals. We derived the conditions under which the complex-conjugate operator of the DAFT, i.e., $\mathbf{A}\mathbf{A}^{\mathsf{T}}$, is dense, and hence, the IQ imbalance may disrupt the inherent sparsity of the AFDM channel matrix. Our analysis additionally showed that residual CFO error resulted in energy leakage in the affine domain. The LMMSE detector was extended to consider the improper Gaussian noise resulting from the IQ imbalance, and its performance approached that of the ideal hardware scenario, similar to the MRC-based DFE detector.


\begin{thebibliography}{00}
\bibitem{bemani2023afdm}
A. Bemani, N. Ksairi, and M. Kountouris, 
``Affine frequency division multiplexing for next generation wireless communications,'' 
\textit{IEEE Transactions on Wireless Communications}, 
vol. 22, no. 11, pp. 8214--8226, Mar. 2023.

\bibitem{wu2024afdmmp}
L. Wu, S. Luo, D. Song, F. Yang, R. Lin, and S. Xie, 
``AFDM signal detection based on message passing scheme,'' 
\textit{Digital Signal Processing}, 
vol. 153, pp. 104633, Jun. 2024.

\bibitem{tao2024afdmim}
Y. Tao, M. Wen, Y. Ge, J. Li, E. Basar, and N. Al-Dhahir,
``Affine frequency division multiplexing with index modulation: Full diversity condition, performance analysis, and low-complexity detection,'' 
\textit{IEEE Journal on Selected Areas in Communications}, vol. 43, no. 4, pp.  1041--1055, Apr. 2025.

\bibitem{xu2024mbuamp}
J. Xu, Z. Liang, and K. Niu, 
``Multi-Block UAMP detection for AFDM under fractional delay-Doppler channel,'' 
\textit{Proceedings of the IEEE Wireless Communication and Networking COnference (WCNC)}, pp. 1--5,
Mar. 2025.

\bibitem{benzine2023afdm}
W. Benzine, A. Bemani, N. Ksairi, and D. Slock, 
``Affine frequency division multiplexing for communications on sparse time-varying channels,'' 
\textit{Proceedings of the IEEE Global Communications Conference (GLOBECOM)}, 
pp. 4921--4926, Dec. 2023.

\bibitem{cao2024afdm}
R. Cao, Y. Zhong, J. Lyu, D. Wang, and L. Fu, 
``AFDM channel estimation in multi-scale multi-lag channels,'' 
\textit{Proceedings of the IEEE Global Communications Conference (GLOBECOM)}, 
pp. 1569--1574, Dec. 2024.

\bibitem{temiz2025afdmplim}
M. Temiz and C. Masouros, 
``Affine frequency division multiplexing with subcarrier power-level index modulation for integrated sensing and communications,'' 
\textit{Proceedings of the IEEE International Workshop on Signal Processing Advances in Wireless Communications (SPAWC)}, 
pp. 1--5, Jul. 2025.

\bibitem{bedeer2025AF}
E. Bedeer, 
``Ambiguity function analysis of affine frequency division multiplexing for ISAC,'' 
\textit{Proceedings of the IEEE Global Communications Conference (GLOBECOM)}, 
pp. 1--5, Dec. 2025.

\bibitem{tandur2007joint}
D. Tandur and M. Moonen, 
``Joint adaptive compensation of transmitter and receiver IQ Imbalance under carrier frequency offset in OFDM-based systems,'' 
\textit{IEEE Transactions on Signal Processing}, 
vol. 55, no. 11, pp. 5246--5252, Nov. 2007.

\bibitem{almradi2015spectral}
A. Almradi and K. A. Hamdi, 
``Spectral efficiency of OFDM systems with random residual CFO,'' 
\textit{IEEE Transactions on Communications}, 
vol. 63, no. 7, pp. 2580--2590, Jul. 2015. 

\bibitem{huemer2013widely}
M. Huemer, A. Onic, C. Hofbauer, and S. Trampitsch, 
``Widely linear data estimation for unique word OFDM,'' 
\textit{Proceedings of the Asilomar Conference on Signals, Systems and Computers}, 
pp. 1934--1938, Nov. 2013. 

\end{thebibliography}
\end{document}